\documentclass[journal]{IEEEtran}

\usepackage{amsmath,amsfonts,amssymb}
\usepackage{graphicx}
\usepackage{xcolor}
\usepackage{multicol}
\usepackage{flushend}

\usepackage{tikz}

\usepackage{pgfplots}
\usepackage{filecontents}
\pgfplotsset{compat=1.5.1}

\tikzset{
  pics/square/.default={1},
  pics/square/.style = {
    code = {
    \draw[pic actions] (0,0) rectangle (#1,0.6#1);
    }
  }
}

\usepackage{bbm}
\newcommand{\IN}{\mathbbm{N}}		
\newcommand{\IR}{\mathbbm{R}}		
\newcommand{\IC}{\mathbbm{C}}		

\renewcommand{\epsilon}{\varepsilon}				
\renewcommand{\theta}{\vartheta}                  	
\renewcommand{\phi}{\varphi}						

\newcommand{\set}[1]{\left\{ #1 \right\}}				
\newcommand{\abs}[1]{\left\lvert #1 \right\rvert}		
\newcommand{\norm}[1]{\left\lVert #1 \right\rVert}		
\newcommand{\onot}[1]{\mathcal{O}{\left( #1 \right)}}	

\newcommand{\signatur}[3]{#1:#2\,\rightarrow\,#3}     

\newcommand{\DF}{\textup{DF}}
\newcommand{\SF}{\textup{SF}}

\newcommand{\Reco}{\textup{Reco}}
\newcommand{\bxi}{ {\boldsymbol\xi} }

\newcommand{\bla}{ {\boldsymbol\lambda} }
\newcommand{\brho}{ {\boldsymbol\rho} }

\newcommand{\zb}[1]{\mbox{\boldmath{${#1}$}}}
\newcommand{\zbs}[1]{\mbox{\boldmath\scriptsize{${#1}$}}}
\newcommand{\zbss}[1]{\mbox{\boldmath\tiny{${#1}$}}}
\renewcommand{\d}{\, \text{d}}
\newcommand{\adj}{{\vdash \hspace*{-1.752mm} \dashv}}

\usepackage[binary-units = true,exponent-product = \cdot]{siunitx}
\DeclareSIUnit{\mup}{\text{$\mu_0$}}

\pgfplotsset{
colormap/inferno/.style={%
    /pgfplots/colormap={inferno}{%
      rgb=(0.001462, 0.000466, 0.013866)
      rgb=(0.037668, 0.025921, 0.132232)
      rgb=(0.116656, 0.047574, 0.272321)
      rgb=(0.217949, 0.036615, 0.383522)
      rgb=(0.316282, 0.053490, 0.425116)
      rgb=(0.410113, 0.087896, 0.433098)
      rgb=(0.503493, 0.121575, 0.423356)
      rgb=(0.596940, 0.154848, 0.398125)
      rgb=(0.688653, 0.192239, 0.357603)
      rgb=(0.775059, 0.239667, 0.303526)
      rgb=(0.851384, 0.302260, 0.239636)
      rgb=(0.912966, 0.381636, 0.169755)
      rgb=(0.956852, 0.475356, 0.094695)
      rgb=(0.981895, 0.579392, 0.026250)
      rgb=(0.987464, 0.690366, 0.079990)
      rgb=(0.973088, 0.805409, 0.216877)
      rgb=(0.947594, 0.917399, 0.410665)
      rgb=(0.988362, 0.998364, 0.644924)
    },
  },}

\definecolor{ibilight}{RGB}{193,216,237}
\definecolor{ibidark}{cmyk}{1.00,0.69,0.00,0.12}
\definecolor{uke2}{cmyk}{0.10,0.18,0.25,0.36}
\definecolor{uke3}{cmyk}{0.00,0.00,0.00,0.80}
\definecolor{ukesec1}{cmyk}{0.00,0.09,1.00,0.00}
\definecolor{ukesec2}{cmyk}{0.00,0.61,0.99,0.00}
\definecolor{ukesec3}{cmyk}{0.59,0.00,0.22,0.00}
\definecolor{ukesec4}{cmyk}{0.54,0.00,1.00,0.00}
\definecolor{tuhh}{cmyk}{0.65,0.00,0.20,0.00}

\usetikzlibrary{positioning}
\usepgfplotslibrary{units}

\usepackage[hyphens]{url}

\begin{document}

\title{Generalized MPI Multi-Patch Reconstruction using Clusters of similar System Matrices}

\author{M.~Boberg, T.~Knopp, P.~Szwargulski, M.~M\"oddel
\thanks{This work was supported by the German Research Foundation (DFG, grant number KN 1108/2-1) and the Federal Ministry of Education and Research (BMBF, grant numbers 05M16GKA and 13XP5060B).}
\thanks{All authors are with the Section for Biomedical Imaging, University Medical Center Hamburg-Eppendorf, Germany and the Institute for Biomedical Imaging, Hamburg University of Technology, Germany (e-mail: m.boberg@uke.de).}
\thanks{Copyright (c) 2019 IEEE. Personal use of this material is permitted. However, permission to use this material for any other purposes must be obtained from the IEEE by sending a request to pubs-permissions@ieee.org.}} 



\maketitle

\begin{abstract}
The tomographic imaging method magnetic particle imaging (MPI) requires a multi-patch approach for capturing large field of views. This approach consists of a continuous or stepwise spatial shift of a small sub-volume of only few cubic centimeters size, which is scanned using one or multiple excitation fields in the kHz range. Under the assumption of ideal magnetic fields, the MPI system matrix is shift invariant and in turn a single matrix suffices for image reconstruction significantly reducing the calibration time and reconstruction effort. 
For large field imperfections, however, the method can lead to severe image artifacts. In the present work we generalize the efficient multi-patch reconstruction to work under non-ideal field conditions, where shift invariance holds only approximately for small shifts of the sub-volume. Patches are clustered based on a magnetic-field-based metric such that in each cluster the shift invariance holds in good approximation. The total number of clusters is the main parameter of our method and allows to trade off calibration time and image artifacts. The magnetic-field-based metric allows to perform the clustering without prior knowledge of the system matrices.
The developed reconstruction algorithm is evaluated on a multi-patch measurement sequence with 15 patches, where efficient multi-patch reconstruction with a single calibration measurement leads to strong image artifacts. Analysis reveals that calibration measurements can be decreased from 15 to 11 with no visible image artifacts. A further reduction to 9 is possible with only slight degradation in image quality.
\end{abstract}

\begin{IEEEkeywords}
Biomedical imaging, focus fields, image reconstruction, magnetic particle imaging
\end{IEEEkeywords}


\section{Introduction}
\IEEEPARstart{T}{he} tomographic imaging technique magnetic particle imaging (MPI)~\cite{knopp2017magnetic} is a promising tool for vascular imaging applications~\cite{vogel2016first}. Diseases like stroke~\cite{ludewig2017magnetic}, stenosis~\cite{vaalma2017magnetic}, and the presence of aneurysms~\cite{sedlacik2016magnetic} can be detected by applying magnetic nanoparticles (MNPs). In addition, MPI allows to visualize coated medical instruments~\cite{haegele2016magnetic} making the technique a promising tool for interventional procedures. 

In MPI different magnetic fields are used to image the distribution of magnetic nanoparticles. Standard MPI systems use one or more dynamic drive fields exciting the magnetization of the nanoparticles and a static selection field spatially encoding the generated magnetization signal. Ideally, the drive fields are realized by homogeneous fields, whereas the selection field is realized by a linear gradient field. The selection field suppresses signal generation outside a small volume around the field free point (FFP) or field free line (FFL), while the drive fields rapidly move this region around. The bulk of the magnetization signal will be generated from locations crossed by the low field region defining the effective field of view of the imaging system. Signal detection is usually done inductively and simplified by using a sinusoidal excitation in the kHz range \cite{knopp2017magnetic}.

The size of the field of view of the imaging sequence above is proportional to the quotient of drive-field amplitude and gradient strength. The two ways to enlarge it are limited in practice. Lowering the gradient strength leads to a loss of resolution~\cite{Rahmer2009BMC}. The drive-field amplitude on the other hand is limited by power loss, tissue heating~\cite{Bohnert2010}, and peripheral nerve stimulation~\cite{Saritas2013TMI,schmale2015mpi}. For human applications the amplitude will be limited to about \SIrange{4}{7}{\milli\tesla\per\mup}~\cite{schmale2015mpi}. Assuming typical gradient strengths of \SIlist{-1;-1;2}{\tesla\per\meter\per\mup} in $x$-, $y$-, and $z$-direction this leads to a field of view of \SI{10 x 10 x 5}{\milli\meter\cubic}. One can increase the effective field of view either by moving the object \cite{szwargulski2018moving} or by using additional low frequency focus fields where the aforementioned restrictions do not apply~\cite{Gleich2010}. These additional fields can be used to continuously or stepwise relocate the FFP or FFL of the selection field and with it the signal generating region. In this work we focus on stepwise focus-field sequences where the focus fields are allowed to change in between excitation cycles only. After each shift a new sub-volume is sampled such that the total field of view can be seen as a patchwork of small sub-volumes, which is why this imaging sequence is referred to as multi-patch sequence~\cite{Knopp2015PhysMedBiol}. 

For the reconstruction of multi-patch MPI data one can perform a patch-wise reconstruction and combine the data in a post-processing step~\cite{ahlborg2016using}. As it has been shown in~\cite{Knopp2015PhysMedBiol} it is advantageous though to apply a joint reconstruction by combining the measured data into a single linear system of equations, which couples the individual sub-volumes, ensures consistency at patch boundaries, and prevents patch boundary related artifacts in the images. However, while the joint reconstruction algorithm leads to very good image quality, its performance is insufficient since calibration time, memory consumption, and runtime performance scale quadratically with the number of patches. A major reduction of the basic approach was achieved in~\cite{szwargulski2018efficient}. In that work ideal magnetic fields were assumed, which allows to relate the system functions of the individual patch measurements via a spatial shift. Exploiting the shift invariance the authors were able to remove the dependence on the number of patches for calibration effort and memory consumption and reduce it to linearity for the runtime of the reconstruction. In practice, field imperfections lead to a violation of the shift invariance. In turn the efficient multi-patch reconstruction causes unwanted reconstruction artifacts~\cite{szwargulski2018efficient}. In the present work we aim to generalize the efficient multi-patch reconstruction to account for field imperfections.

\section{Basic Concept} \label{sec:concept}
\begin{figure*}
    \centering
    \input{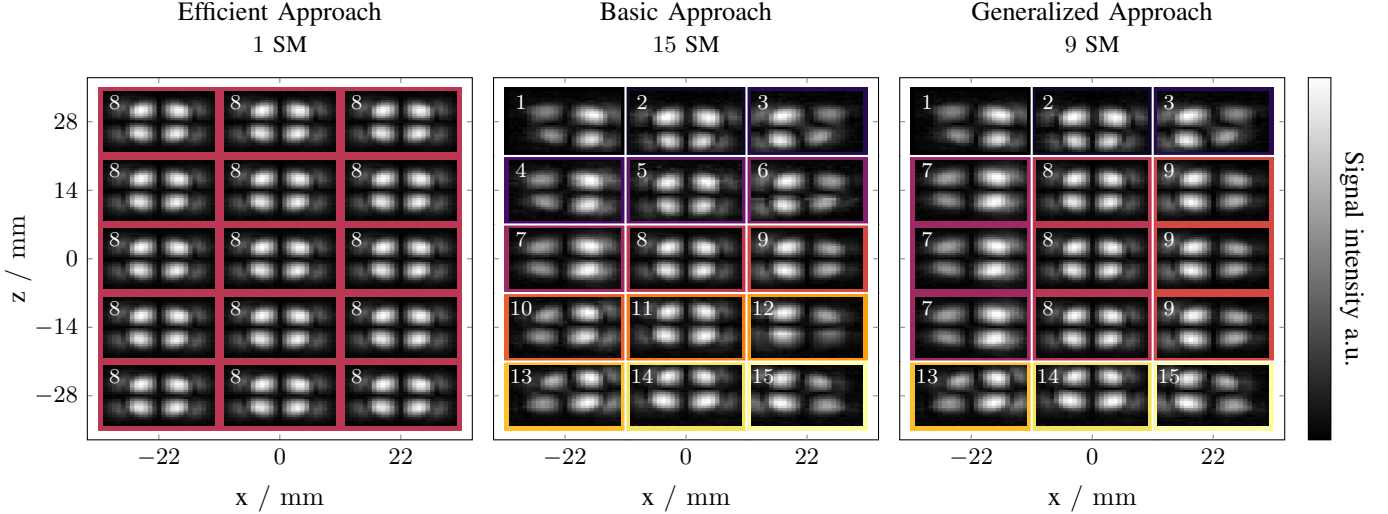}
    \caption{Visualization of the $xz$-slice of the absolute value of the $1733$th system matrix frequency component for each patch of the efficient multi-patch approach~\cite{szwargulski2018efficient} on the left, the basic approach~\cite{Knopp2015PhysMedBiol} in the center and our generalized approach on the right. The system-matrix patch numbers are shown in the upper left corner of each component. For a better visualization of the different approaches each system matrix is marked with an individual color. The spatial structure of the MPI system matrix depends mainly on the magnetic fields~\cite{Rahmer2009BMC}. Therefore, imperfections in the magnetic field of our MPI system~\cite{Weber2016iwmpi} lead to spatial variations in the frequency component for different patches as shown in the center. In the basic approach, these variations are accounted for by obtaining each system matrix individually. The efficient multi-patch reconstruction uses the central system matrix (patch \num{8}) as a replacement to speed up calibration and reconstruction at the cost of reconstruction artifacts. In contrast, our generalized multi-patch reconstruction method uses several system matrices (e.g. \numrange[range-phrase=--]{1}{3}, \numrange[range-phrase=--]{7}{9}, and \numrange[range-phrase=--]{13}{15}) for approximation allowing to trade off between calibration and reconstruction effort on the one hand and image artifacts on the other hand.}
    \label{fig:SM_basic+efficient}
\end{figure*}

To illustrate the main idea of our proposal consider a multi-patch imaging sequence, where a number of sub-volumes cover a larger field of view. For multi-patch reconstructions there are currently two methods available. The basic multi-patch reconstruction~\cite{Knopp2015PhysMedBiol} and the efficient multi-patch reconstruction~\cite{szwargulski2018efficient}. The main difference between both methods is the number of calibration measurements required and the efficiency with which they are used in the reconstruction algorithm. The basic reconstruction requires a calibration measurement for each patch, whereas the efficient method requires one calibration measurement. However, while the former can be always applied, the later can technically only be applied to MPI systems with ideal magnetic fields. In practice, the efficient multi-patch reconstruction will even for imperfect magnetic fields at the expense of reconstruction artifacts. Those artifacts are caused by the fact that the MPI system function is not shift invariant, but varies slightly depending on the patch position as illustrated in Fig.~\ref{fig:SM_basic+efficient}.

With our generalized multi-patch reconstruction approach we aim to account for non-ideal magnetic fields by clustering patches with similar system functions and jointly approximating these system functions by a cluster-specific system function as shown in Fig.~\ref{fig:SM_basic+efficient}. This makes it possible to effectively trade off the calibration effort on the one side with image artifacts on the other. With the number of clusters equal to the number of patches our reconstruction method is an accelerated version of the low artifact and low performance joint multi-patch reconstruction method~\cite{Knopp2015PhysMedBiol}. Meanwhile it is equal to the high performance artifact prone efficient multi-patch reconstruction~\cite{szwargulski2018efficient} for a single cluster. The main challenge of this approach is to find a suitable clustering and cluster-specific system function without any prior-knowledge on calibration measurements. To this end we introduce a magnetic field based error metric, which allows to compare system functions of different patches without measuring them directly. Assuming that the magnetic fields are known, our generalized approach can be broken down into four steps, the first three of which have to be performed only once per multi-patch sequence.
\begin{enumerate}
    \item Choose the number of clusters between one and the number of patches and cluster patches with similar system functions. In this study, we use the k-medoids clustering algorithm to minimize the sum of pairwise errors, where the error of two system functions is measured by the above-mentioned error metric.
    \item Find a cluster-specific system function to approximate the system functions within each cluster as is shown in Fig.~\ref{fig:SM_basic+efficient}. To this end we can directly use the medoids provided by the k-medoids clustering algorithm. Here, a medoid is the system function in a cluster whose total error to all other system functions in the cluster is minimal. We can extend the set of potential cluster-specific system functions beyond those corresponding to patches of the multi-patch imaging sequence and find the one minimizing the total error to all other system functions in the cluster.
    \item Obtain the cluster-specific system functions by performing calibration measurements.
    \item Reconstruct multi-patch data using the cluster-specific system functions and a generalization of the efficient multi-patch reconstruction algorithm.
\end{enumerate}

\section{Continuous Setting} \label{sec:theory}
We consider a multi-patch imaging sequence, where $L \in \IN$ small sub-volumes cover a larger field of view. We let $\bxi_l\in\IR^3$ be the gradient field FFP position of the $l$-th patch and $\Omega_l\subseteq\IR^3$ be the corresponding sub-volume covering all positions where the particles generate a measurable signal upon drive-field excitation. Where a signal is measurable depends on the system function, which falls off with the sensitivity of the receive coil and the potential magnetization dynamics. Magnetic moments far from the volume covered by the FFP trajectory remain in saturation and can only rotate in a small arc, strongly affecting their ability to generate strong signals at higher harmonic frequencies, whereas the signal of the first harmonic is usually superimposed by the feed-through of the excitation signal and hence not recoverable. Therefore, $\Omega_l$ can be chosen slightly larger than the volume covered by the FFP trajectory and assumed to be compact \cite{weber2015artifact}. The domain in which particles generate signal during the multi-patch measurement sequence is then given by
\begin{equation*}
    \Omega = \bigcup_{l \in I_L} \Omega_l,
\end{equation*}
where $I_L := \{ 1,\dots, L \} $. We will not put any major restrictions on the sampled sub-volumes but we assume that the entire sampling volume $\Omega$ is approximately proportional to the number of patches $L$.

The general MPI multi-patch imaging equation is given by the following system of equations
\begin{equation}
	\hat{\zb u}^{\bxi_l} = \int\limits_{\Omega_l}\hat{\zb s}^{\bxi_l}(\zb{r})c(\zb{r})\d\zb r,  \quad l \in I_L,
    \label{eq:MultiPatch}
\end{equation}
where $\hat{\zb u}^{\bxi_l}\in\IC^K$ are the Fourier coefficients of the induced voltage signal measured while sampling $\Omega_l$. We restrict the Fourier coefficients to an upper index $K$ since in practice, the induced voltage signal is sampled with a finite bandwidth. $\signatur{\hat{\zb s}^{\bxi_l}}{\IR^3}{\IC^K} \in \mathfrak{S}$ is the MPI system function where 
\begin{equation*}
\mathfrak{S} := \set{\hat{\zb s}^{\bla} :~\text{gradient field FFP at }\bla\in\IR^3}    
\end{equation*}
is the space of all system functions, which could be part of a multi-patch imaging sequence.  $\signatur{c}{\IR^3}{\IR}$ is the particle concentration that we aim to reconstruct. In the most general setting image reconstruction requires knowledge of all $L$ system functions, which entails a number of time consuming calibration measurements and a memory demanding reconstruction algorithm. To shorten notation we will assume $l\in I_L$ whenever $l$ appears as index. Further, note that we consider multi-patch imaging sequences where the gradient of the selection field remains constant. In this case all potential sub-volume measurements are in one-to-one correspondence to the corresponding FFP position of the gradient field.

\subsection{Approximative Shift Invariance}
\label{sec:shiftinvariance}
Consider a multi-patch setting where we have an ideal linear selection field, perfectly homogeneous excitation fields and homogeneous focus fields, which can be used to shift the FFP of the gradient field away from its central position $\zb{0} \in \IR^3$ to $\bxi_l\in \IR^3$. In this case the system functions are globally shift invariant
\begin{equation}
    \hat{\zb s}^{\bxi_l}(\zb{r}) = \hat{\zb s}^{\zbs{0}}{\left(\zb T^{-\bxi_l}(\zb{r})\right)}. 
    \label{eq:globalshiftinvariance}
\end{equation}
Here, $\hat{\zb s}^{\zbs{0}}$ is the central system function where the focus fields are zero and $\signatur{\zb T^{\bxi}}{\IR^3}{\IR^3}, \zb x \mapsto \zb x + \bxi$ is the spatial translation. Let $\Omega_{\zbs 0}$ be the signal carrying volume of $\hat{\zb s}^{\zbs 0}$. The general MPI multi-patch imaging equation~\eqref{eq:MultiPatch} can be rewritten to contain the central system function only
\begin{equation}
	\hat{\zb u}^{\bxi_l} = \int\limits_{\Omega_{\zbs 0}}\hat{\zb s}^{\zbs{0}}(\zb{r}) c{\left(\zb T^{\bxi_l}(\zb{r})\right)} \d\zb r,
    \label{eq:MultiPatchOrig}
\end{equation}
allowing to decrease reconstruction time, memory consumption and the number of calibration measurements \cite{szwargulski2018efficient}.

\begin{figure}
    \centering
    \begin{tikzpicture}
	\tikzset{filled/.style={fill=gray!50, draw=black},
    point/.style = {draw, circle,  fill = black, inner sep = 1.5pt}}
    
	\def\firstset{plot[smooth cycle] coordinates {(1.5,0.7) (1.9,0.1)  (1.1,-0.2) (-0.3,-1.7) (-1.5,-0.7) (-0.8,0.2) (-1.3,1.0) (-0.2,1.2)}}
	\def\secondset{plot[smooth cycle] coordinates {(0.5,0) (1.5,1.1) (2.5,0.8) (3.4,1.1) (4.0,-0.2) (2.5,-0.5) (1.25,-1.3)}}    
    \def\thirdset{plot[smooth cycle] coordinates {(1.8,-1.3) (2.7,-0.6) (4.1,-0.6) (4.7, 0.1) (4.8,-1.0) (4.0,-1.5)}}
    
    \begin{scope}
        \clip \firstset;
        \fill[filled] \secondset;
    \end{scope}
    \draw \firstset; 
    \draw \secondset;
    \draw \thirdset;

    \node (firstCenter) at (0,0) [point, label = {below: $\bxi_1$}] {};
    \node (secondCenter) at (2.3,0.2) [point, label = {above: $\bxi_2$}] {};
    \node (thirdCenter) at (3.5,-1.1) [point, label = {right: $\bxi_3$}] {};

	\node at (-1.4,-1.3) {$U_1$};
	\node at (3.9,1.2) {$U_2$};
	\node at (5.2,-0.7) {$U_3$};

\end{tikzpicture}
    \caption{Visualization of the sets $U_l$ of three different system functions. Since $U_1$ and $U_2$ intersect it is possible to choose only two system functions with gradient field FFP positions $\bla_1\in U_1\cap U_2$ and $\bla_2\in U_3$ with an error smaller than $\tau$.}
    \label{fig:sets}
\end{figure}
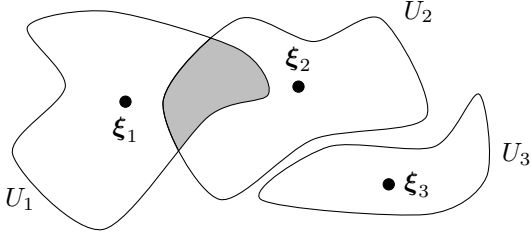

In practice, however, magnetic fields deviate from what they ideally should be, which is why equation~\eqref{eq:globalshiftinvariance} does not hold in this case.  
Therefore, instead of a global shift invariance we can consider local shift invariance where we can find system functions $\hat{\zb s}^{\bla_1}$ and $\hat{\zb s}^{\bla_2}$ with $\bla_1 \neq \bla_2$ satisfying
\begin{equation}
    \hat{\zb s}^{\bla_1}{\left(\zb T^{\bla_1}(\zb{r})\right)} \approx \hat{\zb s}^{\bla_2}{\left(\zb T^{\bla_2}(\zb{r})\right)}\quad \forall \zb r \in \IR^3.
    \label{eq:shiftinvariance}
\end{equation}
Let $\signatur{\delta}{\IC^K\times\IC^K}{\IR^+_0}$ be a semimetric on $\IC^K$ then $\delta$ can be used to quantify the approximation error between the left- and right-hand-side of \eqref{eq:shiftinvariance} for a fixed $\zb r$. Following, we define a semimetric on $\mathfrak S$ by
\begin{align*}
    \begin{split}
        \epsilon: \mathfrak{S}\times\mathfrak{S}\rightarrow &\ \IR^+_0, \\ 
        \left(\hat{\zb s}^{\bla_1},\hat{\zb s}^{\bla_2}\right) \mapsto & \int\limits_{\IR^3} \delta{\left(\hat{\zb s}^{\bla_1}{\left(\zb T^{\bla_1}(\zb r)\right)},\hat{\zb s}^{\bla_2}{\left(\zb T^{\bla_2}(\zb r)\right)}\right)} \d\zb r,
    \end{split}
\end{align*}
which allows to quantify how well $\hat{\zb s}^{\bla_1}$ and $\hat{\zb s}^{\bla_2}$ approximate each other. $\epsilon$ is a semimetric since it holds the identity of indiscernibles and symmetry condition due to the semimetric $\delta$. Based on this metric we define the open subset of gradient field FFP positions 
\begin{equation*}
    U_l := \left\{ \bla \in \IR^3 : \epsilon{\left(\hat{\zb s}^{\bxi_l},\hat{\zb s}^{\bla}\right)} < \tau \right\}
\end{equation*}
for which corresponding system functions $\hat{\zb s}^{\bla}$ approximate $\hat{\zb s}^{\bxi_l}$ with an error below a preset maximal error tolerance $\tau$. Three exemplary sets with corresponding FFP are shown in Fig.~\ref{fig:sets}.

To approximate the $L$
system functions in the signal equation by fewer than $L$ system functions we can use the observation that in cases where multiple $U_l$ intersect the corresponding system functions $\hat{\zb s}^{\bxi_l}$ can be approximated by $\hat{\zb s}^{\bla}$ using equation~\eqref{eq:shiftinvariance}, where $\bla$ is taken from the intersection. In Fig.~\ref{fig:sets}, two system functions with gradient field FFPs $\bla_1\in U_1\cap U_2$ and $\bla_2\in U_3$ can be selected that fulfill the given error tolerance $\tau$. If only one system function is desired, $\tau$ has to be increased until $U_1\cap U_2\cap U_3 \neq \emptyset$.
In particular we propose to find a number of $J \le L$ gradient field FFP positions $\bla_j \in \IR^3, j\in I_J$, and an assignment $\signatur{\iota}{I_L}{I_J}$ satisfying $\bla_{\iota(l)} \in U_{l}$ for all $l\in I_L$, which allows to approximate the signal equation~\eqref{eq:MultiPatch} by
\begin{equation}
	\hat{\zb u}^{\bxi_l} \approx 
	    \int\limits_{\Omega_{\bla_{\iota(l)}}}
	    \hat{\zb s}^{\bla_{\iota(l)}}( \zb{r}) 
	    c{\left(\zb T^{\bxi_l-\bla_{\iota(l)}}(\zb{r})\right)} \d\zb r.
\label{eq:MultiPatchCenterT}
\end{equation}
Here, $\Omega_{\bla_{\iota(l)}}$ is the signal generating sub-volume of $\hat{\zb s}^{\bla_{\iota(l)}}$ and the error tolerance $\tau$ depending on $J$ is chosen sufficiently large to ensure that the gradient field FFP positions and mapping exist. In case of $J < L$ several $l\in I_L$ share the same calibration index $j$, which reduces the number of system functions needed to be known for reconstruction down to $J$. In the extreme case $J=1$ equation~\eqref{eq:MultiPatchCenterT} boils down to equation~\eqref{eq:MultiPatchOrig} when choosing $\bla_1 = \zb{0}$ (cf. Fig~\ref{fig:SM_basic+efficient}).

\subsection{Magnetic-Field-Based Error Metric} \label{sec:selection}
For a given $J$ the selection of the calibration set
\begin{equation*}
    \Lambda = \set{ \bla_j\in\IR^3 : j \in I_J}
\end{equation*}
and the related mapping $\signatur{\iota}{I_L}{I_J}$ satisfying $\bla_{\iota(l)} \in U_l$ for all $l \in I_L$ as proposed in section~\ref{sec:shiftinvariance} will have a significant impact on the reconstruction, as errors introduced by the approximation in equation~\eqref{eq:shiftinvariance} will cause reconstruction artifacts. In order to keep these artifacts low we propose to select $(\Lambda,\iota)$ using an optimization procedure, where the total approximation error is minimized
\begin{equation}
    \underset{(\Lambda,\iota)}{\arg\min}\sum_{l\in I_L} \epsilon{\left(\hat{\zb s}^{\bxi_l},\hat{\zb s}^{\bla_{\iota(l)}}\right)}.
    \label{eq:sfselectionMag}
\end{equation}
$\left(\Lambda,\iota\right)$ can be be seen as a clustering, where $\iota$ maps the $L$ patches to $J$ clusters. All patches in the cluster $j$ share that their respective system function is approximated by the same shifted system function $\hat{\zb s}^{\bla_j}$.

A priori we have no knowledge about any of the system functions. It can only be obtained by calibration measurements. Therefore, an error metric $\varepsilon$ requiring such knowledge would contradict the aim of our proposal to reduce the calibration time and measuring the final selection of system functions only. Here, we propose a more practical approach based on our earlier observation, that deviations in the magnetic fields are the main cause for the approximation error, i.e. to use a magnetic-field-based error metric $\signatur{\mu}{\IR^3\times \IR^3}{\IR^+_0}$, $(\bxi_l,\bla) \mapsto \mu(\bxi_l,\bla)$. We want to replace the comparison of the system functions in each spatial position with $\delta$ by a comparison of the magnetic fields in the same positions. This metric only requires knowledge about the static and dynamic magnetic fields during the measurement and thus can already be applied when planing the calibration measurements. The selection of $(\Lambda,\iota)$ can be analogously performed by
\begin{equation}
    \underset{(\Lambda,\iota)}{\arg\min}\sum_{l\in I_L} \mu{\left(\bxi_l,\bla_{\iota(l)}\right)}.
    \label{eq:sfselectionMagFields}
\end{equation}
Note, however, that in general $\mu(\bxi_l,\bla_1) < \mu(\bxi_l,\bla_2)$ does not imply $\varepsilon{\big(\hat{\zb s}^{\bxi_l},\hat{\zb s}^{\bla_1}\big)} < \varepsilon{\big(\hat{\zb s}^{\bxi_l},\hat{\zb s}^{\bla_2}\big)}$, which is why the optimal tuple $(\Lambda,\iota)$ from equation~\eqref{eq:sfselectionMagFields} will be only suboptimal with respect to the optimization functional in equation~\eqref{eq:sfselectionMag}.

We want to transfer the semimetric $\epsilon$ from the space of system functions to the space of magnetic fields. The magnetic field \mbox{$\signatur{\zb H^\bla}{\IR^3\times\IR}{\IR^3}$} inside the scanner bore is given by
\begin{align}
    \zb H^\bla(\zb r,t) = \zb H^\bla_{\SF}(\zb r) + \sum_{q=1}^Q \zb H^q_{\DF}(\zb r,t),
    \label{eq:magnField}
\end{align}
where $\zb r\in\IR^3$ is the spatial position and $t\in\IR$ is the time. It is the superposition of the selection and focus field $\signatur{\zb H^\bla_{\SF}}{\IR^3}{\IR^3}$ with FFP $\bla\in\IR^3$ and $Q$ drive fields $\signatur{\zb H^q_{\DF}}{\IR^3\times\IR}{\IR^3}$. The drive fields can be separated into the coil sensitivity $\signatur{\zb p^q_{\DF}}{\IR^3}{\IR^3}$ and the current $\signatur{I^q}{\IR}{\IR}$ for $q\in I_Q$. To take different drive-field amplitudes into account we scale the coil sensitivity with the maximum of the applied current $\tilde{\zb H}_\DF^q = \tilde{I}^q\zb p_\DF^q$. It holds that $\zb H^\bla_\SF, \tilde{\zb H}^q_\DF \in C^\infty{\left(\IR^3\right)}:=\set{\signatur{\zb H}{\IR^3}{\IR^3} \text{ smooth}}$ for all $\bla\in\IR^3,q\in I_Q$.
That means we have $Q+1$ static magnetic fields on which the error metric can be built up.

 First, the integration domain of the metric has to be restricted since the magnetic fields do not have a compact support and we are not interested in the entire field. A reasonable restriction is the signal carrying volume $\Omega_\bla$ of the patches, since this volume coincides with the support of the system function $\hat{\zb s}^\bla$. To retain the symmetry of the semimetric we choose $\Omega_{\bla_1,\bla_2} := \zb T^{-\bla_1}{\left(\Omega_{\bla_1}\right)}\,\cup\,\zb T^{-\bla_2}{\left(\Omega_{\bla_2}\right)}$ as the integration domain for the comparison of the magnetic fields of the system functions $\hat{\zb s}^{\bla_1}$ and $\hat{\zb s}^{\bla_2}\in \mathfrak{S}$. The comparison of the magnetic field in each spatial position is done by the semimetric $\delta$, where we use the metric induced by $\norm{\cdot}_2$.

First, we transfer the semimetric $\epsilon$ to the selection and drive fields individually and define
\begin{align*}
    \begin{split}
        &\signatur{\tilde\nu}{\left(C^\infty(\IR^3)\times\IR^3\right)\times \left(C^\infty(\IR^3)\times\IR^3\right)}{\IR_0^+},\\
        &\left(\left(\zb f,\bla_1\right),\left(\zb g,\bla_2\right)\right)\mapsto \negthickspace 
        \int\limits_{\Omega_{\bla_1\,\negthickspace,\bla_2}}\negthickspace\negthickspace {\norm{\zb f{\left(\zb T^{\bla_1}(\zb r)\right)}-\zb g{\left(\zb T^{\bla_2}(\zb r)\right)}}_2} \d\zb r.
    \end{split}
\end{align*}
Then, $\nu_\SF(\bla_1,\bla_2) := \tilde\nu{\big(\big(\zb H^{\bla_1}_\SF,\bla_1\big),\big(\zb H^{\bla_2}_\SF,\bla_2\big)\big)}$ describes the deviation of the selection and focus field whereas $\nu^q_\DF(\bla_1,\bla_2) := \tilde\nu{\big(\big(\tilde{\zb H}^{q}_\DF,\bla_1\big),\big(\tilde{\zb H}^{q}_\DF,\bla_2\big)\big)}$ describes the deviation of the drive fields with respect to the FFPs of the underlying selection field. Note that the abbreviations are only possible if $\bla_1$ and $\bla_2$ are a unique representation of the selection and drive fields. For the metric 
$\signatur{\tilde\mu}{\big(C^\infty(\IR^3)\times\IR^3\big)\times \big(C^\infty(\IR^3)\times\IR^3\big)}{\IR_0^+}$
of the total magnetic field we use the same abbreviation 
\begin{align*}
    \signatur{\mu}{\IR^3\times\IR^3}{\IR^+_0},\,
    \mu(\bla_1,\bla_2) := \tilde\mu{\left(\left(\zb H^1\!,\bla_1\right),\left(\zb H^2\!,\bla_2\right)\right)}.
\end{align*}
According to the superposition in \eqref{eq:magnField} the metric $\mu$ is given by the weighted sum
\begin{align}
    \mu(\bla_1,\bla_2) = 
    \omega_0 \nu_\SF(\bla_1,\bla_2) + \sum_{q=1}^Q \omega_q \nu^q_\DF(\bla_1,\bla_2)
    \label{eq:metric_mu}
\end{align}
with weights $\omega_q\in\IR^+_0$ for $q=0,\dots,Q$, which can be used to balance the contribution of the individual fields to the total error metric.

\section{Discrete Setting} \label{sec:discretization}

The model discussed so-far was discrete in the time respectively frequency dimension and continuous in space. We will next discuss the discretization of space and derive the discrete imaging equation, which will be used for image reconstruction.

\subsection{System Calibration}
We consider a global infinite regular lattice 
\begin{align*}
    \Gamma_\infty &= \left\{ \sum_{i=1}^3 k_i d_i\zb e_i : k_i\in \mathbb Z,\ i=1,2,3 \right\}
\end{align*}
with grid spacing $d_1, d_2, d_3 \in \IR$ and basis $\left\{ \zb e_1, \zb e_2, \zb e_3 \right\} \subseteq \IR^3$ to discretize $\Omega$ by
\begin{equation*}
    \Gamma = \Omega \cap \Gamma_\infty
\end{equation*}
and the $l$-th sub-volume by 
\begin{equation*}
    \Gamma_l = \Omega_l \cap \Gamma_\infty.
\end{equation*}
With $N = \abs{\Gamma}$, $N_l = \abs{\Gamma_l}$ and $\Omega$ being approximately proportional to $L$ we will have that $N$ is approximately proportional to $LN_L$, where $N_L = \max\limits_{l\in I_L} N_l$.

For system calibration, a full drive-field sequence is measured at different calibration sample positions and the resulting data is interpreted as a matrix with the frequency components being the first dimension and the sampling positions being the second. Within \cite{Knopp2015PhysMedBiol} it was proposed to discretize each $\hat{\zb s}^{\bxi_l}$ from equation~\eqref{eq:MultiPatch} on the entire imaging volume. $\hat{\zb s}^{\bxi_l}$ is sampled at $N$ discrete sampling positions $\zb r_n\in\Gamma$. In total this requires $\onot{N_L L^2}$ calibration measurements. In contrast \cite{szwargulski2018efficient} only requires the discretization of the central system function in equation~\eqref{eq:MultiPatchOrig} with sampling positions $\zb r_n\in\Gamma_{\zbs{0}} = \Omega_{\zbs 0}\,\cap\,\Gamma_\infty$. The effort for this procedure is $\onot{N_L}$ calibration measurements. For the discretization of equation~\eqref{eq:MultiPatchCenterT} each of the $J$ system functions has to be discretized. In particular 
\begin{equation*}
    \hat{\zb{S}}^{\bla_{\iota(l)}} := \left( w \hat{s}^{\bla_{\iota(l)}}_{k}(\zb{r}_n) \right)_{k\in I_K; n\in I_{N_{\iota(l)}}}
\end{equation*}
with sampling positions $\zb r_n\in\Gamma_{\iota(l)}$ and weights $w = d_1 d_2 d_3$. Compared to the method proposed in \cite{Knopp2015PhysMedBiol} the sampling domain is restricted to a subset of $\Gamma$ and therefore the calibration effort is reduced to $\onot{N_L J}$.

\subsection{Imaging Equation}
Next, the imaging equations are discretized using the midpoint quadrature rule. Within \cite{Knopp2015PhysMedBiol} it was proposed to discretize equation~\eqref{eq:MultiPatch} on the entire imaging volume $\Omega$. Let $\brho_m$, $m \in I_N$ be the sampling points of the imaging volume then
\begin{equation}
    \hat{u}^{\bxi_l}_{k} = \sum\limits_{m\in I_{N}} \breve{s}_{k,m}^{\bxi_l} c_m,
    \label{eq:multiPatchAll}
\end{equation}
where $\breve{s}_{k,m}^{\bxi_l} := w\hat{s}^{\bxi_l}_{k}(\brho_m)$ are the entries of the system matrix $\hat{\zb{S}}$ and $c_m := c(\brho_m)$ is the discretized particle concentration. This can be expressed in matrix vector notation as
\begin{equation}
    \hat{\zb{S}} \zb{c}=\hat{\zb{u}},
    \label{eq:SCu}
\end{equation}
with particle concentration vector $\zb c := (c_m)_{m\in I_{N}}$ and measurement vector 
\begin{equation*}
	\hat{\zb u}=
	\begin{pmatrix}
		\hat{\zb{u}}^{\bxi_1}\\
		\hat{\zb{u}}^{\bxi_2}\\
		\vdots\\
		\hat{\zb{u}}^{\bxi_L}\\
	\end{pmatrix}\in \IC^{LK}
\end{equation*}
with $\hat{\zb u}^{\bxi_l} := \left(\hat u_k^{\bxi_l}\right)_{k\in I_K}$.

In contrast, we discretize the integral in \eqref{eq:MultiPatchCenterT} similarly to the method proposed in~\cite{szwargulski2018efficient}. To this end we use the individual system matrix grids $\zb r_n\in\Gamma_{\iota(l)}$, $n \in I_{N_{\iota(l)}}$ to discretize equation~\eqref{eq:MultiPatchCenterT} by 
\begin{equation*}
    \hat{u}^{\bxi_l}_{k} = \sum\limits_{n\in I_{N_{\iota(l)}}} \hat{s}_{k,n}^{\bla_{\iota(l)}} c^l_n,
\end{equation*}
where $\hat{s}_{k,n}^{\bla_{\iota(l)}}$ are the entries of the system matrix $\hat{\zb S}^{\bla_{\iota(l)}}$ and $c^l_n := c{\big(\zb T^{\bxi_l-\bla_{\iota(l)}}(\zb r_n)\big)}$. In order to relate $c^l_n$ to the global discretization of the particle concentration, we assume that $\zb T^{\bxi_l-\bla_{\iota(l)}}{(\Gamma_{\iota(l)})} \subseteq \Gamma$ for all $l \in I_L$. This allows us to implicitly define an index mapping $\phi^l:I_{N_{\iota(l)}} \rightarrow I_{N}$ from the elements of $\Gamma_{\iota(l)}$ to the elements of $\Gamma$ by
\begin{equation*}
    \brho_{\phi^l(n)} = \zb T^{\bxi_l-\bla_{\iota(l)}}(\zb r_n).
\end{equation*}
The imaging equation can then be written as
\begin{equation}
    \hat{u}^{\bxi_l}_{k} = \sum\limits_{n\in I_{N_{\iota(l)}}}\hat{s}_{k,n}^{\bla_{\iota(l)}} c_{\phi^l(n)},
    \label{eq:multiPatchDiscrete2}
\end{equation}
where $ c_{\phi^l(n)} := c{\big(\brho_{\phi^l(n)}\big)}$. Equation~\eqref{eq:multiPatchDiscrete2} can also be brought into matrix vector notation~\eqref{eq:SCu} by expressing the system matrix by
\begin{equation*}
    \hat{\zb S}=
    \begin{pmatrix}
        \hat{\zb{S}}^{\bla_{\iota(1)}} \zb P^1\\
        \hat{\zb{S}}^{\bla_{\iota(2)}} \zb P^2\\
        \vdots\\
        \hat{\zb{S}}^{\bla_{\iota(L)}} \zb P^L
    \end{pmatrix} \in \IC^{LK\times N}
\end{equation*}
where 
$
\zb P^l := \left( \delta_{\phi^l(n),m} \right)_{n \in I_{N_{\iota(l)}}, m \in I_{N}}.
$

\subsection{Adjoint Imaging Equation}
The imaging equation \eqref{eq:multiPatchDiscrete2} describes a space- and time-efficient way to carry out matrix-vector multiplications with the matrix $\hat{\zb S}$. Many iterative solvers additionally require the multiplication with the adjoint $\hat{\zb S}^\adj$, i.e. $\zb x = \hat{\zb S}^\adj \zb y$ with
$\zb x = \left( x_{m}\right)_{m \in N}$ and $\zb y = \left( \hat{y}^{l}_{k} \right)_{l \in I_L, k \in I_K}$. We first consider again the dense representation of the system matrix from \cite{Knopp2015PhysMedBiol} for which  $\zb x = \hat{\zb S}^\adj \zb y$ can be expressed as
\begin{equation*}
    x_m = \sum\limits_{l \in I_L} \sum\limits_{k \in I_K}  \overline{\breve{s}_{k,m}^{\bxi_l}} y^{l}_{k}, \quad m\in I_{N}.
\end{equation*}
In order to express the summation in terms of the calibration scans for our proposed approach we first define the index set
\begin{equation*}
    \Theta^m = \set{ (l,n) \in I_L\times\IN  : n \in I_{N_{\iota(l)}} \land  \phi^l(n) = m }
\end{equation*}
that represents for each image position $\brho_m$ in the global grid $\Gamma$ the set of patch indices for which $\brho_m = \zb r_n \in \Gamma_l$. With that, the adjoint imaging equation can be expressed as
\begin{equation}
    x_m = \sum\limits_{(l,n) \in \Theta^m} \sum\limits_{k \in I_K}  \overline{ \hat{s}_{k,n}^{\bla_{\iota(l)}} } y^{l}_{k}, \quad m\in I_{N}.
    \label{eq:multiPatchDiscreteTransp}
\end{equation}
Here, we note that the set $\Theta^m$ contains only a single index when the grids $\Gamma_l$ are disjoint. In practice the system matrix grids will slightly overlap because of the overscan that is usually used when acquiring the system matrix \cite{weber2015artifact}. It might also happen that there is an $\tilde{m} \in I_N$ for which $\Theta^{\tilde{m}} = \emptyset$ in which case $x_m$ would be set to zero. In a practical implementation one will not iterate over the index $I_{N}$ but instead will iterate over all $I_{N_{\iota(l)}}$ for $l\in I_L$ in which case the outer sum on the right hand side of the equation has always at least one summand. When iterating over the individual subgrids, one also does not need to arrange the set $\Theta^m$ explicitly but one can initialize the vector $\zb x$ to zero and add the result of the inner sum to those regions in $\zb x$ being effected by the considered patch.

\subsection{Image Reconstruction}
Most MPI publications treat \eqref{eq:SCu} as an inverse problem that is solved using a regularized least squares optimization approach
\begin{equation} \label{Eq:LS}
    \zb c_\text{Reco} = \underset{\boldsymbol{c}\in\IR^{N}\!,\boldsymbol{c}\geq 0}{\arg\!\min} \| \hat{\zb{S}}\zb{c}-\hat{\zb{u}} \|^2_2 + R(\zb c)
\end{equation}
where $R$ is a regularization term that can for instance be the L$_2$ norm $\norm{ \cdot }_2$. Equation~\eqref{Eq:LS} is usually solved using iterative solvers like Krylov subspace methods (e.g. conjugate gradient least square method), row- or column-action methods (such as the Kaczmarz method), or splitting methods (like the alternating direction method of multipliers) that allow for incorporating sophisticated regularizers \cite{storath2017edge}.

The Krylov subspace methods and the splitting methods have in common that they apply in each iteration multiplications with the system matrix $\hat{\zb S}$ and its adjoint $\hat{\zb S}^\adj$. The multiplication with $\hat{\zb S}$ can be carried out very efficiently by evaluating \eqref{eq:multiPatchDiscrete2} instead of \eqref{eq:multiPatchAll} while the multiplication with the adjoint matrix can be carried out using \eqref{eq:multiPatchDiscreteTransp}. 

The row-action Kaczmarz method is very popular within MPI since it shows rapid convergence, which is based on the high orthogonality of the system matrix rows in MPI \cite{Knopp2010PhysMedBio}. The Kaczmarz method applies in each iteration inner products between a certain matrix row $\hat{\zb s}_k^{\bla_{\iota(l)}} $ of $\hat{\zb S}^{\bla_{\iota(l)}}$ and some temporary vector $\zb x$. Those inner products can be carried out the same way as \eqref{eq:multiPatchDiscrete2} whereas the second operation in the Kaczmarz algorithm is a vector update $\zb d:=\zb d_\text{old} + \alpha\hat{\zb s}_k^{\bla_{\iota(l)}} $ that can also be carried out by looping over the system matrix grid $\Gamma_{\iota(l)}$ only.

\subsection{Discrete Error Metric}\label{ssec:DiscretErrorMetric}
For the discretization of the error metric we also use the midpoint quadrature rule. Therefore, a discretization of the integration domain $\Omega_{\bla_1,\bla_2}$ is required. We use the sampling points of the central system matrix $\Gamma_{\zbs{0}}$ for each metric. This leads to 
\begin{align*}
    \nu_\SF{\left(\bla_1\,\negthickspace,\,\negthickspace\bla_2\right)} \,\negthickspace
    =   \,\negthickspace
        \frac{1}{N_0}\!\sum_{\zbs r_n\in\Gamma_{\zbss{0}}}\negthickspace\,\negthickspace
        \norm{\zb H^{\bla_1}_{\SF}\!{\left(\zb T^{\bla_1}(\zb r_n)\,\negthickspace\right)}
        \!-\,\negthickspace\zb H_{\SF}^{\bla_2}\!{\left(\zb T^{\bla_2}(\zb r_n)\,\negthickspace\right)}}_2
\end{align*}
for the selection field and
\begin{align*}
    \nu_\DF^q{\left(\bla_1\,\negthickspace,\,\negthickspace\bla_2\right)} \,\negthickspace
    =   \,\negthickspace
        \frac{1}{N_0}\!\sum_{\zbs r_n\in\Gamma_{\zbss{0}}}\negthickspace\,\negthickspace
        \norm{\tilde{\zb H}^q_{\DF}{\left(\zb T^{\bla_1}(\zb r_n)\,\negthickspace\right)}
        \!-\,\negthickspace\tilde{\zb H}^q_{\DF}{\left(\zb T^{\bla_2}(\zb r_n)\negmedspace\right)}\!}_2
\end{align*}
for all drive-field coil sensitivities $q\in I_Q$.
Combining these semimetrics just like in \eqref{eq:metric_mu} leads to a discretized error metric on the total magnetic field.

\section{Methods} \label{sec:methods}

All experiments in this work were performed using a pre-clinical MPI scanner (Bruker, Ettlingen) that is equipped with a 3D drive-field generator, a 3D focus-field generator and selection-field generator orientated in vertical direction ($z$-direction). The selection-field gradient was set to \SI{-0.75}{\tesla\per\meter\per\mup} in $x$- and $y$-direction and \SI{1.5}{\tesla\per\meter\per\mup} in $z$-direction. The drive fields had a frequency of $f_x = \frac{2.5}{102}\si{\mega\hertz}$, $f_y = \frac{2.5}{96}\si{\mega\hertz}$, and $f_z = \frac{2.5}{99}\si{\mega\hertz}$ resulting in a period length of $T_{\DF} \approx \SI{21.5}{\milli\second}$. The amplitudes of all three drive fields were set to $A_x=A_y=A_z=\SI{12}{\milli\tesla\per\mup}$ resulting in a field of view of size \SI{32.0 x 32.0 x 16.0}{\milli\meter}. The focus fields can be adjusted between \SIlist{-17;17}{\milli\tesla\per\mup} in $x$- and $y$-direction and \SIlist{-42;42}{\milli\tesla\per\mup} in $z$-direction.

For the object measurements we applied a multi-patch sequence where $L=15$ patches were sequentially measured. At each patch position \num{100} drive-field cycles were measured requiring about $\SI{2.154}{\second}$ pure measurement time. Changing the focus fields requires \num{7} drive-field cycles until the field reaches its final value. Thus, the \num{15} patches required in total a measurement time of about $\SI{34.42}{\second}$. The \num{15} patches were arranged on a $3\times 5$ grid within the $xz$-plane of the scanner. Within the $x$-direction the position $\bxi$ was shifted to $\xi_x \in \set{-22,0,22}\si{\milli\meter}$. The five positions in $z$-direction were chosen as $\xi_z \in \set{-28,-14,0,14,28}\si{\milli\meter}$. Additionally, two system matrices were measured at $\xi_x\in\set{-22,22}\si{\milli\meter}$ and $\xi_z = \SI{-22}{\milli\meter}$ for further improvements.

We measured system matrices at all patch positions using a delta sample of size \SI{2x2x1}{\milli\meter} filled with diluted (concentration \SI{250}{\milli\mole\of{Fe}\per\liter}) ferucarbotran (Resovist, I'rom Pharmaceuticals, Tokyo, Japan). Each system matrix was measured at $25 \times 21 \times 27 = 14175$ positions covering a signal carrying volume of \SI{50x42x27}{\milli\meter}. The center of the grid was adjusted to the respective focus-field shift $\bxi_l$ for $l\in I_L$. Each system matrix has a size of $14175\times 80787$ where $80787$ is the product of $K=26929$ frequency components and three receive channels. Each system matrix thus requires \SI{17.5}{\giga\bel} of main memory when storing the complex data in double precision floating point format. The acquisition time for one system matrix was \num{8}\;hours \num{39}\;minutes and \num{27}\;seconds using an averaging factor of \num{50} and \num{190} background measurements. In total, the system matrix acquisition for all $15$ patches required \num{6}\;days.

\begin{figure}
    \centering
    \input{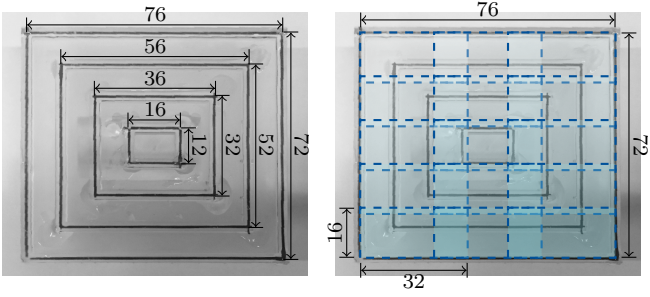}
    \caption{A photo of our multi-patch measurement phantom is shown on the left hand side with an overlay of the $15$ patches on the right. Phantom and patch sizes are given in \si{\milli\meter}. The phantom consists of four concentrically aligned rectangles. Each rectangle has a square internal cross-section of \SI{1x1}{\milli\meter} and is filled with diluted ferucarbotran with a concentration of \SI{250}{\milli\mole\of{Fe}\per\liter}.}
    \label{Fig:Phantom}
\end{figure}

For object measurements we used a 3D printed phantom consisting of four square-shaped nested tubes in the $xz$-plane (see Fig.~\ref{Fig:Phantom}). Each tube has a square cross section of size \SI{1x1}{\milli\meter}. The squares range from \SI{16 x 12}{\milli\meter} to \SI{76x72}{\milli\meter} in \SI{20}{\milli\meter} steps. The purpose of this phantom is to visualize artifacts due to field imperfections that occur when perfect fields are assumed \cite{szwargulski2018efficient}. 

For image reconstruction we chose a grid of size $49\times 21\times 86$ covering a volume of \SI{98x42x86}{\milli\meter}. We used a regularized form \cite{Knopp2010PhysMedBio} of the Kaczmarz algorithm applying \num{3} iterations and a relative regularization parameter of $\lambda_\text{rel} = 0.01$. Only frequencies above \SI{60}{\kilo\hertz} with a signal-to-noise ratio above \num{10} were used for reconstruction to remove most of the non-static background signal~\cite{them2016sensitivity}. This reduced the number of frequency components to \num{1956}. Those parameters provided a good balance between a high spatial resolution and a high image signal-to-noise ratio.

All algorithms were implemented in the programming language Julia (version 1.2) and the reconstruction algorithm developed in section~\ref{sec:discretization} was integrated into the open-source project \texttt{MPIReco.jl} (version 0.1.1) \cite{MPIReco}.

\subsection{Magnetic Field Representation}
We consider that each static magnetic field is given as solid harmonic expansion 
\begin{align}
    \zb H^\bla_{\SF}(\zb r) &= \sum_{\kappa=0}^{\infty}\sum_{\eta=-\kappa}^{\kappa} \boldsymbol\gamma^\bla_{\kappa,\eta} R_{\kappa}^{\eta}(\zb r)
    \label{eq:HSF_expansion}\\
    \tilde{\zb H}^q_{\DF}(\zb r) &= \tilde{I}^q\sum_{\kappa=0}^{\infty}\sum_{\eta=-\kappa}^{\kappa} \boldsymbol\beta^q_{\kappa,\eta} R_{\kappa}^{\eta}(\zb r)
    \label{eq:pDF_expansion}
\end{align} 
with coefficients $\boldsymbol\gamma^\bla_{\kappa,\eta}, \boldsymbol\beta^q_{\kappa,\eta} \in \IR^3$ and normalized solid harmonics $\signatur{R_{\kappa}^{\eta}}{\IR^3}{\IR}$ as introduced in \cite{Weber2016iwmpi}, \cite{Boberg2017}. 
We determined the coefficients of the selection fields by measuring a spherical $8$-design on the surface of a ball $\mathcal{B}\subseteq\IR^3$ with radius \SI{42}{\milli\meter} around the FFP using a Hall-effect sensor, which took about \SI{45}{\minute}. Taking into account the finite size of the Hall probe and a safety margin, the sphere within the scanner bore with a radius of \SI{59.5}{\milli\meter} was selected as large as possible. The coil sensitivities of the drive fields were simulated with the Biot-Savart law at the same positions. This yields accurate coefficients up to degree $4$ by an equally weighted quadrature \cite{Beentjes2015}, \cite{Hardin1996}.
We note that \eqref{eq:HSF_expansion} and \eqref{eq:pDF_expansion} are restricted to the ball $\mathcal{B}$ and it holds that $\zb H^\bla_\SF, \tilde{\zb H}^q_\DF \in C^\infty{\left(\mathcal{B}\right)}$ for all $\bla\in\mathcal{B},q\in I_Q$.

 Therefore, it must hold for the error metric $\mu$ that $\zb T^{\bla_i}{\big(\Omega_{\bla_1,\bla_2}\big)} \subseteq \mathcal{B}$ for $i=1,2$ and all $\bla$ that are relevant for a possible reconstruction to gain accurate results.
The weights of the error metric are chosen to be 
\begin{align*}
    \omega_0 &= \frac{1}{\max\limits_{\zbs r_n\in\Gamma_{\zbss{0}}, l\in I_L}\norm{\zb H_{\SF}^{\bxi_l}\left(\zb T^{\bxi_l}(\zb r_n)\right)}_2}\\
    \omega_q &= \frac{1}{\max\limits_{\zbs r_n\in\Gamma_{\zbss{0}}, l\in I_L}\norm{\tilde{\zb H}^q_{\DF}\left(\zb T^{\bxi_l}(\zb r_n)\right)}_2},\quad q=1,2,3,
\end{align*}
which leads to a normalization of the contributing fields. Since we use the same amplitude for each drive field the maximum current is cancelled out by the weights and the metric only depends on the coil sensitivity of the drive fields.

\subsection{Clustering}
With the error metric, we have a basis to find the optimal tuple $(\Lambda,\iota)$. For tackling the optimization problem, we use the clustering algorithm k-medoids \cite{Clustering}. We will first consider a clustering where the FFP positions for the system matrices are a subset of the FFP positions $\bxi_l$ used during the measurement. In this case the clustering groups the given patches into $J$ number of clusters based on the cost matrix $\left(\mu(\bxi_l,\bxi_j)\right)_{l,j\in I_L}$. The algorithm returns the set $\Lambda\subseteq\set{\bxi_l : l\in I_L}$, $\abs{\Lambda}=J$, and the mapping $\signatur{\iota}{I_L}{I_J}$ that leads to the smallest total cost given by 
$$ \sum_{l\in I_L} \mu{\left(\bxi_l,\bxi_{\iota(l)}\right)}.$$ 
For this form of clustering we always have patches where system matrices are directly available and we have patches where the system matrix is approximated. 

In the next step we consider a less constrained approach. Instead of $\Lambda \subseteq \set{\bxi_l : l\in I_L}$ we consider $\Lambda \subseteq \Gamma$, i.e. we allow the calibration FFP positions to lie on any point of the reconstruction grid $\Gamma$. The set $\Lambda$ thus contains elements $\bla_j, j\in I_J$ and we seek for an appropriated $\Lambda$ and $\iota$ that maps from the measured FFP positions to the calibration FFP positions. Since the simultaneous optimization of $\Lambda$ and $\iota$ (i.e. \eqref{eq:sfselectionMagFields}) is a computationally intensive task, we do not tackle this optimization problem directly but instead use a two step procedure. First, $\iota$ is determined by restricting the positions to $\set{\bxi_l : l\in I_L}$. Then we optimize for each cluster $j\in I_J$ the functional
\begin{align*}
    \min_{\bla\in\Gamma} \sum_{\substack{l\in I_L\\ \text{ with } \iota(l) = j}} \mu{\left(\bxi_l,\bla\right)}.
\end{align*}
Since $\set{\bxi_l : l\in I_L} \subseteq \Gamma$, the calculated solution is at least as good as the one that was restricted to the measurement FFP positions.

\section{Results} \label{sec:results}
\subsection{Reconstruction and Clustering}

\begin{figure*}
    \centering
    \input{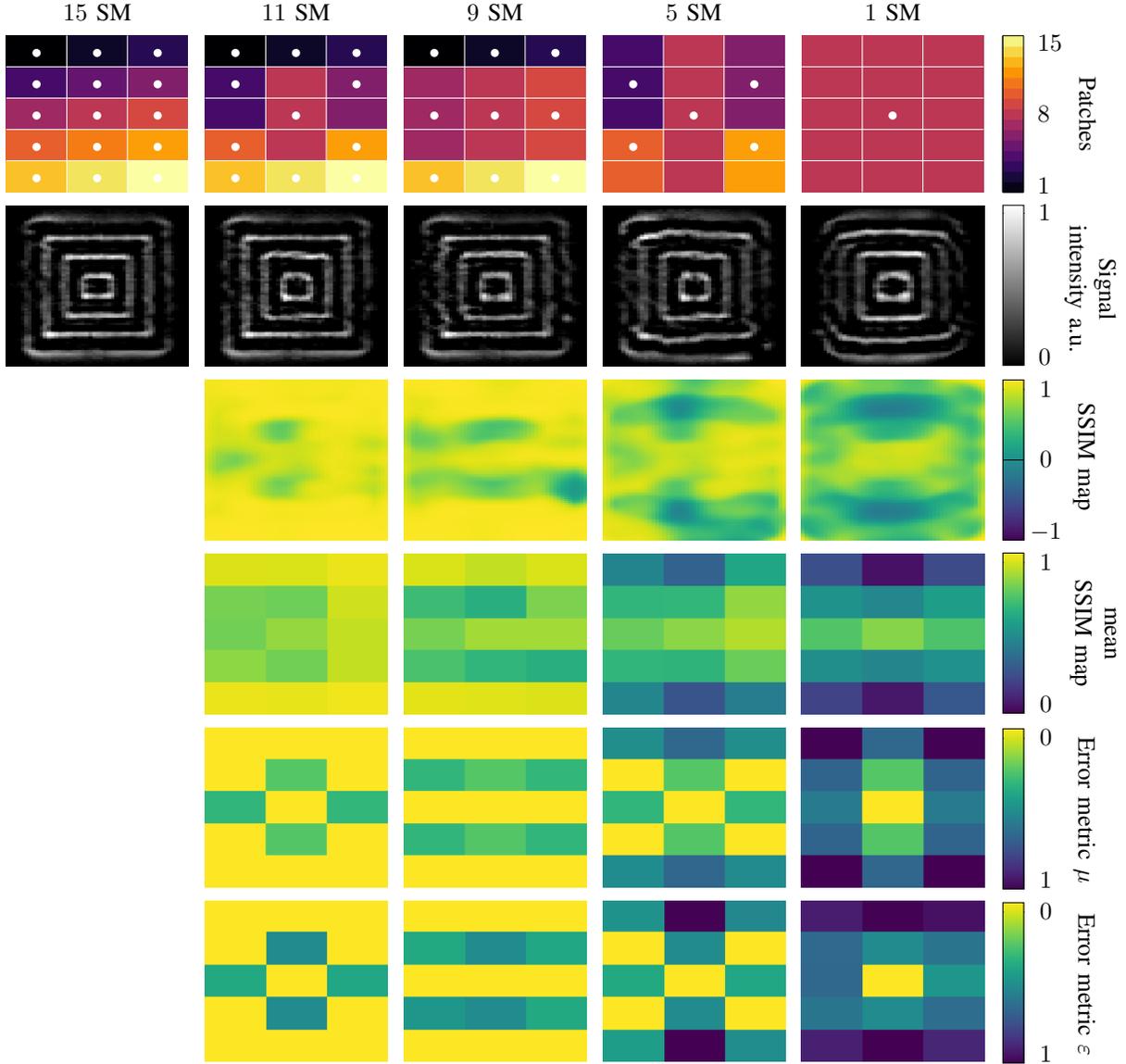}
    \caption{Cluster results for different numbers of system matrices. 
    Shown are the selected system matrices and cluster (first row), the reconstructed images (second row), and the SSIM map compared to the reconstruction with all $15$ system matrices (third row). In both cases the $12$th $xz$-slice of the 3D reconstruction volume is shown. It should be noted that the optically better appearance of the horizontal lines compared to the vertical lines in the reconstructed images is due to the anisotropic resolution, which is only half as high in the $x$-direction as in the $z$-direction. The selected matrices are indicated with a white dot and range from all matrices (first column) until only one matrix (last column). The colormap indicates for every patch which system matrix is used. In the last two rows, the error metrics $\mu{\big(\bxi_l,\bxi_{\iota(l)}\big)}$ and $\epsilon{\big(\hat{\zb s}^{\bxi_l},\hat{\zb s}^{\bxi_{\iota(l)}}\big)}$ for each patch $l\in I_L$ of the selections are shown. For a better comparison, the mean SSIM index for each patch in the reconstructed image is shown in the fourth row.}
    \label{fig:RecoUKE}
\end{figure*}

The cluster and reconstruction results for the phantom are shown in the first rows in Fig.~\ref{fig:RecoUKE}. For the generalized multi-patch approach $5, 9$, and $11$ system matrices are chosen, which can be compared to the basic approach \cite{Knopp2015PhysMedBiol} with $15$ system matrices on the left and the efficient approach \cite{szwargulski2018efficient} with the central system matrix on the right. In the first row, the selected calibration scans are visualized and indicated with a white dot. The colormap encodes for each patch the selected system matrix within a cluster. Below, the \num{12}-th $xz$-slice of the corresponding reconstructed image is shown.  Below the reconstructed images one can find the structured similarity (SSIM) map between the reconstruction result that uses all $15$ system matrices ($\zb c_\text{Reco}^{15}$) and the corresponding dataset in the respective column ($\zb c_\text{Reco}^l$ for $l\in \{11,9,5,1\}$). The SSIM index varies between $-1$ and $1$ where $1$ indicates perfect similarity and $-1$ indicates highest dissimilarity \cite{Brunet2012}.

For comparison, the $12$-th $xz$-slice of the $1733$th frequency component for all $15$ system matrices is shown in Fig.~\ref{fig:SM_basic+efficient} in the middle. The clustering coincides with the visual impact. 
Neighboring patches are combined since they show high similarity. As an example, the chosen system matrices for $J=9$ are shown on the right in Fig.~\ref{fig:SM_basic+efficient} where one can see only small deviations compared to the originally measured $15$ system matrices.

When comparing the reconstruction results $\zb c_\text{Reco}^{15}$ and $\zb c_\text{Reco}^{11}$ one can hardly see any difference. The SSIM index of $\zb c_{\Reco}^{11}$ compared to the reconstruction with all $15$ system matrices is $0.892$. Also, the reconstruction result with nine system matrices looks very similar with an SSIM index of $0.837$. The differences are visible in the vertical edges, especially on the left side. By using only five system matrices the SSIM index drops to $0.699$. Now even the horizontal edges are no longer straight lines. In the reconstruction result with only the central system matrix these lines are more straight again but the two outer rectangles are not connected anymore in the corners. This leads to an SSIM index of $0.591$. Especially for nine and one system matrices the error map of the SSIM index shows well, which system matrices were used for the reconstruction since the SSIM index in the other patches is lower.

\subsection{Error Metric}
In the fifth row of Fig.~\ref{fig:RecoUKE}, the error metric $\mu{\big(\bxi_l,\bxi_{\iota(l)}\big)}$ of each patch $l$ is shown. For a better comparison, the mean SSIM index for each patch is shown in the row above. The structure of the deviations of the patches is similar in each column. But there are also some inconsistencies visible. Due to the overlap of the system matrices the deviations of the reconstructed images are distributed over neighboring patches. Therefore, also the patches whose system matrices are used for reconstruction have a mean SSIM index smaller than $1$. This is not captured by the error metric. Additionally, for the case of $J=1$, the error distribution in the first and last row of the patches is different. While the SSIM based on the reconstruction result shows a convex behavior, the error $\mu$ is concave along the horizontal axis. We note that those differences are not unexpected since the SSIM is highly object specific while the error metric $\mu$ takes the entire field of view into account.

The error metric $\mu$ is transferred from the error metric $\epsilon$ on the system functions to the underlying magnetic fields. For a comparison of both metrics, the metric $\epsilon$ was discretized as it was done for $\mu$ in section~\ref{ssec:DiscretErrorMetric} and $\delta(\zb a,\zb b)$ was chosen to be $\norm{\zb a - \zb b}_2$ for $\zb a, \zb b\in \IC^K$. In the last line in Fig.~\ref{fig:RecoUKE}, it is visualized for each patch. Both metrics show a very similar error distribution, which shows that the field-based error metric is suitable for replacing the system-function-based metric, which would usually not be present when planning the calibration scans. Just for the first and last row the central patch again shows larger deviations for the field-based metric than for the system-function-based metric.

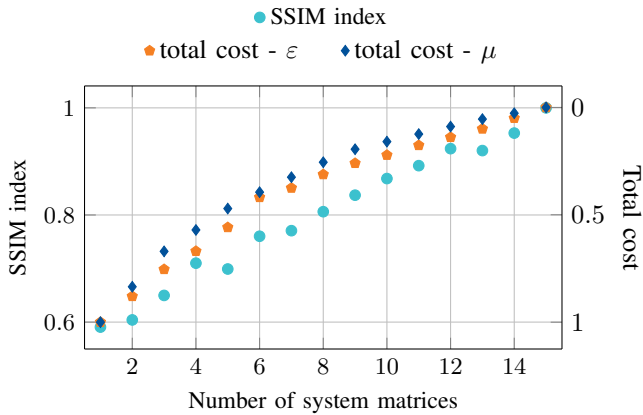
\begin{figure}
    \centering
    \begin{tikzpicture}
\begin{axis}[scale only axis,
    xmin=0.5,
    xmax=15.5,
    width=0.35\textwidth,
    height=3.5cm,
    grid=major,
    only marks,
    xlabel={Number of system matrices},
    ylabel={SSIM index},
    extra x tick labels={},
    extra y tick labels={},
    extra x tick style={major grid style={thin,gray}},
    extra y tick style={major grid style={thin,gray}},
    axis y line*=left,
    style = {font=\small},
    legend style={at={(0.5, 1.18)},anchor = south,draw=none,}
    ]
    \addplot[tuhh,mark=*] table [x index = {0}, y index = {1}, col sep=comma] {tikz/bober5a.csv};
    \addlegendentry{SSIM index}
\end{axis}

\begin{axis}[ scale only axis, 
    xmin=0.5,xmax=15.5, 
    ymin=-0.1,ymax=1.125,
    axis y line*=right, 
    axis x line=none,
    width=0.35\textwidth,
    height=3.5cm,
    only marks, 
    y dir=reverse,
    ylabel={Total cost},
    ylabel style = {rotate=180,yshift=0cm},
    legend style={at={(0.5, 1.05)},anchor = south,draw=none,
    				/tikz/every even column/.append style={column sep=0.5cm}},
                    legend columns=2,
    ]
    \addplot[ukesec2,mark=pentagon*] table [x index = {0}, y index = {3}, col sep=comma] {tikz/bober5a.csv}; 
    \addlegendentry{total cost - $\epsilon$};
    \addplot[ibidark,mark=diamond*] table [x index = {0}, y index = {2}, col sep=comma] {tikz/bober5a.csv}; 
    \addlegendentry{total cost - $\mu$};
\end{axis}
\end{tikzpicture}
    \caption{The SSIM index quantifies the similarity of the optimal reconstruction using $15$ system matrices with multi-patch reconstructions using $15$ to $1$ system matrices. It is compared to the corresponding normalized total costs calculated by the metrics $\epsilon$, based on the system matrices, and $\mu$ based on the magnetic field.}
    \label{fig:SSIM}
\end{figure}

To capture the global progress of the different similarity measures, we plot the SSIM index on the entire grid, the total error $\epsilon$, and the total error $\mu$ in Fig.~\ref{fig:SSIM}. With the exception of a normalization factor the SSIM index for the whole reconstructed image and the total costs for both error metrics shows the same behavior regarding the number of system matrices used for reconstruction. As expected the total costs and overall also the SSIM index grow with increasing number of system matrices. When comparing $\mu$ and $\epsilon$ one can see that the total cost is nearly the same. This underlines that the magnetic fields are a suitable choice for characterizing imperfections in the system matrices prior to calibration.

\subsection{Improved Cluster Positions}
\begin{figure}
    \centering
    \input{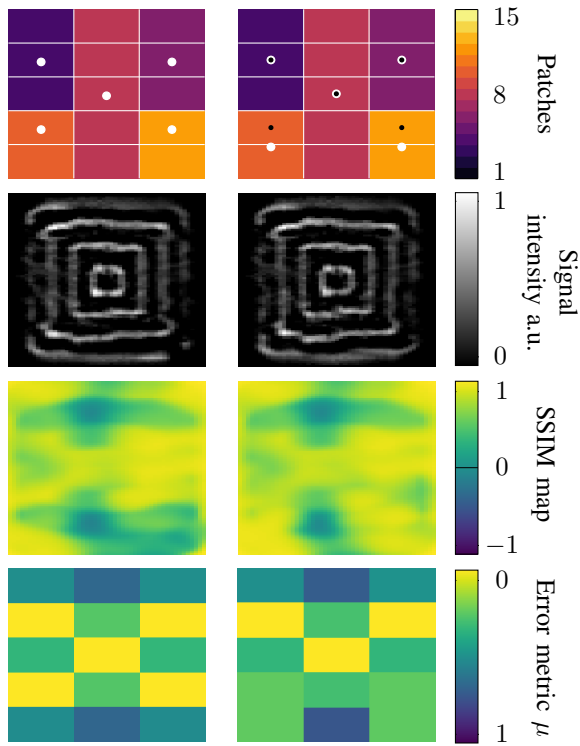}
    \caption{Improved cluster positions for $5$ system matrices. In the left column the basic cluster from Fig.~\ref{fig:RecoUKE} is shown while in the right column the results for the same cluster with improved center positions are illustrated. The first row shows the clusters with the selected matrices, which are indicated with a white dot. The smaller black dot in the right image indicates the system matrices from basic cluster which is also the basis for the colormap. It indicates for each patch, which system matrix is used. Below, the corresponding reconstruction results are shown with the SSIM map compared to the reconstructed image with all $15$ system matrices. In the last row, the error metric $\mu{\left(\bxi_l,\bla_{\iota(l)}\right)}$ for each patch $l\in I_L$ is shown.}
    \label{fig:improvedCluster}
\end{figure}

Until now we considered the clustering where the system matrix FFP positions were a subset of the FFP positions during the measurement. When allowing to take an arbitrary position on the grid spanned over a particular cluster, it is potentially possible to decrease the overall error. The results of the improved clustering is shown in Fig.~\ref{fig:improvedCluster} for $J=5$ compared to the first results from Fig.~\ref{fig:RecoUKE}. In the first row, the clusters and calibration positions are shown. As a reference, the calibration positions of the first clustering are indicated with a small black dot. New calibration positions occur only if the cluster is not symmetric around one patch. Hence, the new calibration positions ($\xi_z = \SI{-22}{\milli\meter}$) are just between the two patches for both asymmetric clusters ($\xi_z \in \set{-14,-28}\si{\milli\meter}$). In the second and third row, the reconstructed image and corresponding SSIM map compared to the reconstruction with all $15$ system matrices is shown. While the SSIM map shows noticeable image enhancements in the lower left corner, it also shows minor image degradation in other areas such as the patch above. The overall SSIM index is slightly improved using the improved cluster positions. The reason for this is that global vertically mispositioning of the horizontal structure at the lower edge of the image, which is less pronounced with improved clustering, affects the SSIM index much more than local variations along the structure. We note that the mispositioning is only improved in the lower left patch while the lower middle patch is not improved since the same system matrix is taken. Finally, Fig.~\ref{fig:improvedCluster} shows the results of the error metric $\mu$ for the improved cluster in the last row. The adapted calibration positions lead to a wider distribution of the errors in the corresponding patches. This reduces the error of each patch and has a similar behavior as the SSIM map.

\subsection{Reconstruction Times}
\begin{figure}
    \centering
    \begin{tikzpicture}
\begin{axis}[ scale only axis, 
    xmin=0.5,xmax=15.5, 
    width=7cm,
    height=3cm,
    only marks, 
    grid,
    unit markings=slash space,
    ylabel={Time},
    y unit = s,
    xlabel={Number of system matrices},
    legend cell align=left,
    legend style={at={(0.5, 1.05)},anchor = south,draw=none,
    				/tikz/every even column/.append style={column sep=0.5cm}},
                    legend columns=1, align=left,
    ]
    \addplot[ibidark,mark=diamond*] table [x index = {0}, y index = {1}, col sep=comma] {tikz/bober7a.csv};
    \addlegendentry{total reconstruction time};
    \addplot[ukesec2,mark=pentagon*] table [x index = {0}, y index = {2}, col sep=comma] {tikz/bober7a.csv};
    \addlegendentry{time per iteration};
\end{axis}
\end{tikzpicture}
    \caption{Total reconstruction times include three Kaczmarz iterations. They are shown in blue depending on the number of system matrices used. In comparison, the times per Kaczmarz iteration are shown in orange.}
    \label{fig:recoTimes}
\end{figure}
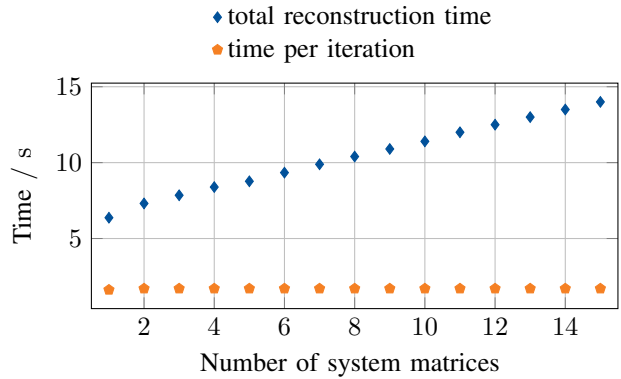

To investigate the reconstruction time we performed benchmarks on a workstation equipped with two Intel Xeon CPU E5-2640 v3 CPUs running at \SI{2.6}{\giga\hertz} and a main memory of \SI{512}{\giga\byte}. Each reconstruction is performed on a single thread ten times and the shortest time of the benchmark series is counted. The total reconstruction times and time per Kaczmarz iteration are shown in Fig.~\ref{fig:recoTimes}. 

Theoretically, the reconstruction time would not depend on the number of calibrations scans since the sparsity of the system matrix is not effected by changing the number of calibration scans. However, when looking at the total reconstruction time one can still see a linear dependence. The reason is that the total reconstruction time includes the time that is required to load the system matrices. The more system matrices have to be loaded, the higher is the total reconstruction time. When looking at the time per iteration, it is nearly constant for $2$ to $15$ system matrices, which fulfills the expectations. For a single patch the reconstruction time is slightly improved from \SI{1.70}{\second} ($2$ patches) to \SI{1.62}{\second} ($1$ patch). One explanation for this reduction in reconstruction time is that a larger number of system matrices in memory leads to more CPU cache misses, which in turn shortens the processing time of the matrix vector operations involved in the Kaczmarz algorithm.

For comparison we also performed a reconstruction where the full multi-patch system matrix is explicitly arranged \cite{Knopp2015PhysMedBiol}. Since the calibration data is only available at subsets of the full reconstruction grid we had to zero-pad the matrix. The total reconstruction using the explicitly arranged system matrix took \SI{58.60}{\second} while generating the same image as the proposed algorithm when using all $15$ patches.

\section{Discussion} \label{sec:discussion}

The proposed multi-patch reconstruction algorithm provides a flexible platform for choosing the calibration scan positions in magnetic particle imaging. In terms of image quality our new algorithm allows to trade off image quality against calibration time by choosing the number of calibration measurements. For the shown dataset we were able to reduce the number of calibration scans from \num{15} to \num{11} marking a reduction of about \SI{26}{\percent} in calibration time while the SSIM index is about $0.89$. A reduction to \num{9} calibration scans (reduction of \SI{40}{\percent} in calibration time) already showed some visible artifacts that might not be acceptable in practice, which is captured by an SSIM index of $0.84$. Reductions to \num{5} or even just \num{1} calibration scan lead to even larger artifacts that should be avoided if it is feasible to spend the time for the calibration scans. Our results indicate that as a rule of thumb increasing the number of patches improves imaging performance as measured by the SSIM index shown in Fig.~\ref{fig:SSIM}. However, there are some exceptions to this rule, e.g. increasing $J$ from $4$ to $5$ and from $12$ to $13$ yields a slight drop in imaging performance. Hence, a prediction of the imaging performance as a function of number of patches is only  possible in good approximation.

Our work is based on \cite{szwargulski2018efficient} where field imperfections were entirely neglected. Our algorithm is a direct generalization where \cite{szwargulski2018efficient} was restricted to $J=1$ and $\bla_1 = \zb 0$. The strength of the generalized algorithm is that it strongly increases the flexibility for the operator of an MPI scanner. One might start with a central system matrix to perform initial multi-patch imaging and accept the image artifacts during animal experiments. After the experiments, one can then reduce the artifact level by acquiring additional system matrices in off-center positions.

One alternative is to plan the number of calibration scans prior to the experiments. We introduced an error metric to determine the deviations in the system functions. While it is simple to define the error metric on the system functions we instead proposed a method that exploits differences in the underlying magnetic fields. This has the huge advantage that the metric can be calculated prior to system calibration. The proposed error metric depends on the selection and focus field, the drive-field coil sensitivities, and the drive-field amplitudes in each patch with the advantage that selection and focus field and drive-field coil sensitivities only need to be measured once for a specific scanner. Moreover, the metric is independent of the particle magnetization dynamics and time evolution of the excitation fields. MPI trajectories influence the metric directly via the maximum drive-field currents and indirectly via the integration domain regardless of whether they are one-, two- or three-dimensional. The same holds true for the selection field, which is explicitly evaluated and indirectly influences the metric via the integration domain. With our choice of weights the metric is invariant under simultaneous upscaling of drive and selection fields, which would leave the patch positions of a multi-patch sequence unchanged.

We noticed that the resulting cluster coincide with the visual impression of the system matrices. The error metric also agrees well with the reconstruction result, where indeed the error was marginal when using a high number of calibration scans while the error was high when using only few calibrations scans. In practice, this implies that one already knows prior to the experiment, which system matrices should be acquired, and which can be neglected. The discussed reconstruction does not make any assumptions despite approximate shift invariance and thus can be used in a very flexible manner. Our framework is even flexible enough to apply it to multi-gradient imaging sequences, where the different patches have been measured with different resolutions respectively gradient strengths \cite{gdaniec2017fast}. 

\begin{table}
    \renewcommand{\arraystretch}{1.2}
    \setlength{\tabcolsep}{5pt}
    \caption{Comparison of Different Multi-Patch Reconstruction Algorithms}
    \label{tab:timesMPapproaches}
    \centering
    \begin{tabular}{c|c|c|c}
        &calibration time & reconstruction time & memory consumpt.\\ \hline
        \cite{Knopp2015PhysMedBiol} & $N_L L^2$ & $IKN_L L^2$ & $KN_L L^2$  \\\hline
        \cite{szwargulski2018efficient} & $N_L $ & $IKN_L L$ & $KN_L $  \\\hline
        proposal & $N_L J$ & $IKN_L L$ & $KN_L J$ 
    \end{tabular}
    \renewcommand{\arraystretch}{1}
\end{table}

For the analysis of the algorithmic complexity we consider three different categories: calibration time, reconstruction time, and memory consumption. Table~\ref{tab:timesMPapproaches} compares these categories for the algorithms developed in \cite{Knopp2015PhysMedBiol} and \cite{szwargulski2018efficient} with our proposed algorithm.
One can see, \cite{szwargulski2018efficient} and our method have an advantage of $L$ in runtime speed over \cite{Knopp2015PhysMedBiol}.  
The only disadvantage is an increase in calibration time and memory consumption compared to \cite{szwargulski2018efficient} because more system matrices are used. Hence, one has to find a balance between calibration time respectively memory consumption and the impact of magnetic field imperfections on the reconstructed image.  
For the concrete setup of \num{15} patches considered in this work, the saving in calibration time was a factor of $6.2$ compared to \cite{Knopp2015PhysMedBiol} and we were able to acquire the data in \num{6} days instead of \num{34} days assuming that the system matrices are measured \num{24} hours per day. Due to the overlap of the system matrix grids the time saving is less then a factor of \num{15}. When decreasing the number of clusters from $J=15$ to $J=11$ or $J=9$ the calibration time can be further reduced to less than $4$ days. 

While the reduction from \num{15} calibration scans to \num{11} calibration scans is just a moderate saving in calibration time, there is still potential room for further improvements. On the one hand, we observed that some of the system matrices were slightly scaled in space. One might be able to introduce a rigid or even non-rigid transformation to cope for these effects and in turn allow to reduce the number of necessary calibration scans even further. An alternative is to exploit symmetries as it has been discussed in \cite{gruttner2013system}. In both cases our reconstruction framework can be used almost unchanged. One general question will then be if the necessary transformations (shifting and/or mirroring) should be done prior to the reconstruction which implies manifolding the amount of system matrices kept in memory during reconstruction, or if the transformations should be done on the fly during reconstruction. The former solution can be seen as a cached version of the latter and in turn one will have to trade off time complexity versus space complexity here.

\bibliographystyle{IEEEtran}

\bibliography{ref}

\begin{thebibliography}{10}
\providecommand{\url}[1]{#1}
\csname url@samestyle\endcsname
\providecommand{\newblock}{\relax}
\providecommand{\bibinfo}[2]{#2}
\providecommand{\BIBentrySTDinterwordspacing}{\spaceskip=0pt\relax}
\providecommand{\BIBentryALTinterwordstretchfactor}{4}
\providecommand{\BIBentryALTinterwordspacing}{\spaceskip=\fontdimen2\font plus
\BIBentryALTinterwordstretchfactor\fontdimen3\font minus
  \fontdimen4\font\relax}
\providecommand{\BIBforeignlanguage}[2]{{%
\expandafter\ifx\csname l@#1\endcsname\relax
\typeout{** WARNING: IEEEtran.bst: No hyphenation pattern has been}%
\typeout{** loaded for the language `#1'. Using the pattern for}%
\typeout{** the default language instead.}%
\else
\language=\csname l@#1\endcsname
\fi
#2}}
\providecommand{\BIBdecl}{\relax}
\BIBdecl

\bibitem{knopp2017magnetic}
T.~Knopp, N.~Gdaniec, and M.~M{\"o}ddel, ``Magnetic particle imaging: from
  proof of principle to preclinical applications,'' \emph{Physics in Medicine
  \& Biology}, vol.~62, no.~14, p. R124, 2017.

\bibitem{vogel2016first}
P.~Vogel, M.~R{\"u}ckert, P.~Klauer, W.~Kullmann, P.~Jakob, and V.~Behr,
  ``First in vivo traveling wave magnetic particle imaging of a beating mouse
  heart,'' \emph{Physics in Medicine \& Biology}, vol.~61, no.~18, p. 6620,
  2016.

\bibitem{ludewig2017magnetic}
P.~Ludewig, N.~Gdaniec, J.~Sedlacik, N.~D. Forkert, P.~Szwargulski, M.~Graeser
  \emph{et~al.}, ``Magnetic particle imaging for real-time perfusion imaging in
  acute stroke,'' \emph{ACS nano}, vol.~11, no.~10, pp. 10\,480--10\,488, 2017.

\bibitem{vaalma2017magnetic}
S.~Vaalma, J.~Rahmer, N.~Panagiotopoulos, R.~L. Duschka, J.~Borgert,
  J.~Barkhausen \emph{et~al.}, ``Magnetic particle imaging ({MPI}):
  Experimental quantification of vascular stenosis using stationary stenosis
  phantoms,'' \emph{PloS one}, vol.~12, no.~1, p. e0168902, 2017.

\bibitem{sedlacik2016magnetic}
J.~Sedlacik, A.~Fr{\"o}lich, J.~Spallek, N.~D. Forkert, T.~D. Faizy, F.~Werner
  \emph{et~al.}, ``Magnetic particle imaging for high temporal resolution
  assessment of aneurysm hemodynamics,'' \emph{PloS one}, vol.~11, no.~8, p.
  e0160097, 2016.

\bibitem{haegele2016magnetic}
J.~Haegele, N.~Panagiotopoulos, S.~Cremers, J.~Rahmer, J.~Franke, R.~L. Duschka
  \emph{et~al.}, ``Magnetic particle imaging: A resovist based marking
  technology for guide wires and catheters for vascular interventions,''
  \emph{IEEE transactions on medical imaging}, vol.~35, no.~10, pp. 2312--2318,
  2016.

\bibitem{Rahmer2009BMC}
J.~Rahmer, J.~Weizenecker, B.~Gleich, and J.~Borgert, ``Signal encoding in
  magnetic particle imaging,'' \emph{BMC Medical Imaging}, vol.~9, no.~4, 2009.

\bibitem{Bohnert2010}
J.~Bohnert and O.~D{\"o}ssel, ``\BIBforeignlanguage{english}{Calculation and
  evaluation of current densities and thermal heating in the body during
  {MPI}}.''\hskip 1em plus 0.5em minus 0.4em\relax {World Scientific,
  Singapore}, 2010, pp. 162--168.

\bibitem{Saritas2013TMI}
E.~U. Saritas, P.~W. Goodwill, G.~Z. Zhang, and S.~M. Conoll,
  ``Magnetostimulation limits in magnetic particle imaging,'' \emph{IEEE
  Transactions on Medical Imaging}, vol.~32, no.~9, pp. 1600 -- 1610, 2013.

\bibitem{schmale2015mpi}
I.~Schmale, B.~Gleich, J.~Rahmer, C.~Bontus, J.~Schmidt, and J.~Borgert,
  ``{MPI} safety in the view of {MRI} safety standards,'' \emph{IEEE
  Transactions on Magnetics}, vol.~51, no.~2, pp. 1--4, 2015.

\bibitem{szwargulski2018moving}
P.~Szwargulski, N.~Gdaniec, M.~Graeser, M.~M{\"o}ddel, F.~Griese, K.~M.
  Krishnan \emph{et~al.}, ``Moving table magnetic particle imaging: a stepwise
  approach preserving high spatio-temporal resolution,'' \emph{Journal of
  Medical Imaging}, vol.~5, no.~4, p. 046002, 2018.

\bibitem{Gleich2010}
B.~Gleich, J.~Weizenecker, H.~Timminger, C.~Bontus, I.~Schmale, J.~Rahmer
  \emph{et~al.}, ``Fast {MPI} demonstrator with enlarged field of view,'' in
  \emph{Proc. ISMRM}, vol.~18, Stockholm, Mai 2010, p. 218.

\bibitem{Knopp2015PhysMedBiol}
T.~Knopp, K.~Them, M.~Kaul, and N.~Gdaniec, ``Joint reconstruction of
  non-overlapping magnetic particle imaging focus-field data,'' \emph{Physics
  in Medicine and Biology}, vol.~60, p. L15, 2015.

\bibitem{ahlborg2016using}
M.~Ahlborg, C.~Kaethner, T.~Knopp, P.~Szwargulski, and T.~Buzug, ``Using data
  redundancy gained by patch overlaps to reduce truncation artifacts in
  magnetic particle imaging,'' \emph{Physics in Medicine and Biology}, vol.~61,
  no.~12, pp. 4583--4598, 2016.

\bibitem{szwargulski2018efficient}
P.~Szwargulski, M.~M{\"o}ddel, N.~Gdaniec, and T.~Knopp, ``Efficient joint
  image reconstruction of multi-patch data reusing a single system matrix in
  magnetic particle imaging,'' \emph{IEEE transactions on medical imaging},
  2018.

\bibitem{Weber2016iwmpi}
A.~Weber, J.~Weizenecker, R.~Pietig, U.~Heinen, and T.~M. Buzug, ``Controlling
  the position of the field-free-point in magnetic particle imaging,''
  \emph{Book of Abstracts IWMPI}, 2016.

\bibitem{weber2015artifact}
A.~Weber, F.~Werner, J.~Weizenecker, T.~Buzug, and T.~Knopp, ``Artifact free
  reconstruction with the system matrix approach by overscanning the
  field-free-point trajectory in magnetic particle imaging,'' \emph{Physics in
  Medicine \& Biology}, vol.~61, no.~2, p. 475, 2015.

\bibitem{storath2017edge}
M.~Storath, C.~Brandt, M.~Hofmann, T.~Knopp, J.~Salamon, A.~Weber, and
  A.~Weinmann, ``Edge preserving and noise reducing reconstruction for magnetic
  particle imaging.'' \emph{IEEE Trans. Med. Imaging}, vol.~36, no.~1, pp.
  74--85, 2017.

\bibitem{Knopp2010PhysMedBio}
T.~Knopp, J.~Rahmer, T.~F. Sattel, S.~Biederer, J.~Weizenecker, B.~Gleich
  \emph{et~al.}, ``Weighted iterative reconstruction for magnetic particle
  imaging,'' \emph{Physics in Medicine and Biology}, vol.~55, no.~6, pp. 1577
  -- 1589, 2010.

\bibitem{them2016sensitivity}
K.~Them, M.~G. Kaul, C.~Jung, M.~Hofmann, T.~Mummert, F.~Werner, and T.~Knopp,
  ``Sensitivity enhancement in magnetic particle imaging by background
  subtraction,'' \emph{IEEE transactions on medical imaging}, vol.~35, no.~3,
  pp. 893--900, 2016.

\bibitem{MPIReco}
T.~Knopp, P.~Szwargulski, F.~Griese, M.~Grosser, M.~Boberg, and M.~M\"oddel,
  ``{MPIReco.jl}: Julia package for image reconstruction in {MPI},''
  \emph{International Journal on Magnetic Particle Imaging}, vol.~4, no.~2,
  2019.

\bibitem{Boberg2017}
M.~Boberg, T.~Knopp, and M.~M\"oddel, ``Analysis and comparison of magnetic
  fields in {MPI} using spherical harmonic expansions,'' \emph{Book of
  Abstracts IWMPI}, 2017.

\bibitem{Beentjes2015}
C.~H.~L. Beentjes, ``Quadrature on a spherical surface,'' \emph{Working note
  available on the website http://people.maths.ox.ac.uk/beentjes/Essays}, 2015.

\bibitem{Hardin1996}
R.~H. Hardin and N.~J.~A. Sloane, ``Mclaren's improved snub cube and other new
  spherical designs in three dimensions,'' \emph{Discrete Comput. Geom.},
  vol.~15, no.~4, pp. 429--441, Apr 1996.

\bibitem{Clustering}
``Clustering.jl: Julia package for data clustering,''
  \url{https://github.com/JuliaStats/Clustering.jl}, accessed: 2019-02-28.

\bibitem{Brunet2012}
D.~Brunet, E.~R. Vrscay, and Z.~Wang, ``On the mathematical properties of the
  structural similarity index,'' \emph{IEEE Transactions on Image Processing},
  vol.~21, no.~4, pp. 1488--1499, 2012.

\bibitem{gdaniec2017fast}
N.~Gdaniec, P.~Szwargulski, and T.~Knopp, ``Fast multiresolution data
  acquisition for magnetic particle imaging using adaptive feature detection,''
  \emph{Medical physics}, vol.~44, no.~12, pp. 6456--6460, 2017.

\bibitem{gruttner2013system}
M.~Gr{\"u}ttner, T.~F. Sattel, F.~Griese, and T.~M. Buzug, ``System matrices
  for field of view patches in magnetic particle imaging,'' in \emph{Medical
  Imaging 2013: Biomedical Applications in Molecular, Structural, and
  Functional Imaging}, vol. 8672.\hskip 1em plus 0.5em minus 0.4em\relax
  International Society for Optics and Photonics, 2013, p. 86721A.

\end{thebibliography}




\end{document}